# Strategies for zero boil-off liquid hydrogen transfer: an export terminal case-study

Halvor Aarnes Krog[1], David Berstad[2].

[1] SINTEF Energy Research, Trondheim, Norway
(e-mail: halvor.krog@sintef.no).

[2] SINTEF Nordvest, Ålesund, Norway

**Abstract**: To ensure economic viability, $LH_2$ export terminals must minimize boil-off losses. We show two strategies to achieve zero boil-off losses for the transfer of 160 000 $m^3$ $LH_2$ (11 248 tons) using a centrifugal pump. In the first strategy, a pump with variable speed drive (VSD) and split-range control for the flow rate achieves losses from 0 wt% to 0.24 wt% in an uncertainty analysis. A pump efficiency approaching 70% is the most important factor to minimize losses. In contrast, a fixed-speed pump has unacceptably high losses ranging from 0.76 wt% to 1.06 wt% (119 tons per ship). The second strategy is to increase the maximum pressure in the seaborne tank (base case is 1.15 bara). Zero loss is achieved for the fixed speed pump if the maximum pressure is increased to 1.35 bara, while 1.22 bara is required for the pump with VSD assuming an efficiency of 60%.

## 1 INTRODUCTION

The Hydrogen Council projects that 200 million tons per annum (MTPA) of clean hydrogen (both low carbon and renewable) requires long-distance transportation in 2050, with 10% (20 MTPA) shipped by sea [1]. Another techno-economic analysis suggests that 84% of the green hydrogen produced in 2050 will require intercontinental maritime transport, and liquid hydrogen ($LH_2$) is one of the most likely transport mediums [2]. Although $LH_2$ shipping is not yet commercialized, it has been demonstrated in pilot-scale by Suiso Frontier (1250 $m^3$). Furthermore, Kawasaki Heavy Industries have an approval-in-principle (AiP) for a commercial 160 000 $m^3$ $LH_2$ carrier [3]. Likewise, HD KSOE's 80 000 $m^3$ $LH_2$ carrier design has received AiP [4]. However, to realize economically viable $LH_2$ shipping, ground infrastructure at the ports must also be optimized and qualified. Boil-off gas (BOG) management is important in this regard.

At export terminals, BOG is formed in the onshore storage tanks and during the transfer process between the onshore tank and the seaborne tanks (ship tanks). Minimizing the amount of BOG during transfer is important for several reasons. First, there will be less gas to reliquefy. This saves OPEX, and it may allow using smaller (and cheaper) equipment when designing the plant (e.g. pipes, valves and possibly the liquefaction plant itself). Second, the BOG may be re-injected into the liquefier, still at cryogenic temperature, by an ejector if the BOG flowrate is stable and sufficiently small compared to the flowrate of "fresh" $H_2$ feed entering the liquefier [5]. Thus, the need for dedicated and expensive BOG compressors may be omitted.

There are two options for $LH_2$ transfer between tanks. The most energetically efficient method is to use a pump. However, the most common method today is the pressure differential method, where some $LH_2$ in the source tank is vaporized to raise the tank pressure above the receiving tank's pressure. Gil-Esmendia et. al. [6] compared the two transfer methods and found that the relative loss (vented gas relative to the transferred $LH_2$ amount) was significantly reduced when using a pump (0%-16% loss) compared to the pressure differential method (20% loss). The lowest loss when using the pump was for a slow filling rate and low initial pressure in the receiving tank. Their case study was however for relatively small tanks (maximum 18 $m^3$ tanks) with high maximum working pressure (12 bara). This is significantly different from the conditions for a prospective $LH_2$ carrier, where each tank is orders of magnitudes larger and the maximum working pressure is typically in the range 1 atm to 2 atm [7]. Furthermore, they modelled an idealized pump without any performance curves.

The importance of the receiving tank's pressure is also noted in Shell's technical magazine [5] in a study on $LH_2$ export terminals. Here, increasing the maximum pressure by 0.1 bar in the seaborne tank reduced the BOG flowrate to the liquefier by approximately 25%. Unfortunately, few details about their method were provided. Similarly, Petitpas [8] found that the losses varied from 2% to 18% when varying the initial pressure in the receiving tank in simulations of an $LH_2$ trailer ($\approx$ 60 $m^3$) unloading to a storage tank using the pressure differential method.

The literature review shows that the pressure in the seaborne tank is a critical parameter, and that a slow filling reduces the BOG. However, to the best of our knowledge, the implications of the process conditions along the $LH_2$ transfer path have not been analyzed previously. This is relevant since we can alter the process conditions by using different control strategies.

We address this research gap by investigating LH$_2$ transfer using a fixed-speed submersible centrifugal pump and a submersible centrifugal pump with a variable speed drive (VSD). The pump with VSD allows to reduce the pump's discharge pressure compared to the fixed-speed pump, and we show that the VSD significantly reduces the exergy destruction of the transfer process. Furthermore, we find that a slow filling rate does indeed reduce the BOG amounts, *but this is only true for the pump with the VSD*. For a fixed-speed centrifugal pump, a slow filling rate increases the loss.

This reveals a second research gap, namely the need for a more comprehensive sensitivity analysis for LH$_2$ transfer systems. To address this, we conduct an uncertainty and global sensitivity analysis (U&GSA) for the pump with VSD where we consider 6 uncertain parameters related to the pump, tanks and pipeline. In the U&GSA the maximum pressure in the seaborne tank is fixed since i) it has a dominating influence on the results, and ii) we conduct a separate analysis where only the maximum pressure in the seaborne tank is varied for both the fixed-speed pump and the pump with VSD.

The article is structured in the following way. Section 2 describes the case study. Section 3 contains the following results. Section 3.1 contains all results where the maximum pressure in the seaborne tank is fixed. This includes the loss when varying filling rate only, a comparison of the exergy destruction for the case of a fixed-speed pump and a pump with VSD, and the U&GSA for a pump with VSD (with a fixed maximum pressure in the seaborne tank). Section 3.2 compares the loss when varying the maximum pressure in the seaborne tank for both pump configurations. Section 4 contains a discussion before concluding in Section 5.

## 2   MATERIALS AND METHODS

The objective is to transfer a total of 160 000 m$^3$ LH$_2$ at a steady-state loading rate of 13 000 m$^3$/h. To this end, we consider four independent loading trains, where each train transfers LH$_2$ from one onshore tank to one seaborne tank. Figure 1 shows the model for one train. Each train has two LH$_2$ and vapor return pipes, but only one is shown for brevity. Note that BOG from the ship is returned to the onshore tank by free flow. As a simplification, the onshore tank does not receive LH$_2$ from the liquefaction plant during the bunkering operation. The model is implemented in Dymola using the TIL library [9].

### 2.1   Control structure

We will consider two variations of flow control where the manipulated variables are:

- Throttling valve opening and pump speed (split-range control).

- Only throttling valve opening (fixed speed pump).

Figure 1 shows a split-range controller [10] where a VSD is in installed on the pump's motor. For the fixed-speed pump, there is no line from the flow controller (FC) to the pump in Figure 1. The split-range control works as follows. For low capacities, the pump runs at minimum speed and the flow rate is controlled by the valve. For high capacities, the valve is fully open and flow is controlled by the pump speed. The benefit of this configuration is that throttling losses are minimized. The drawback is that VSDs are expensive and can introduce vibration-related issues.

All PID controllers are tuned by the SIMC rules [11].

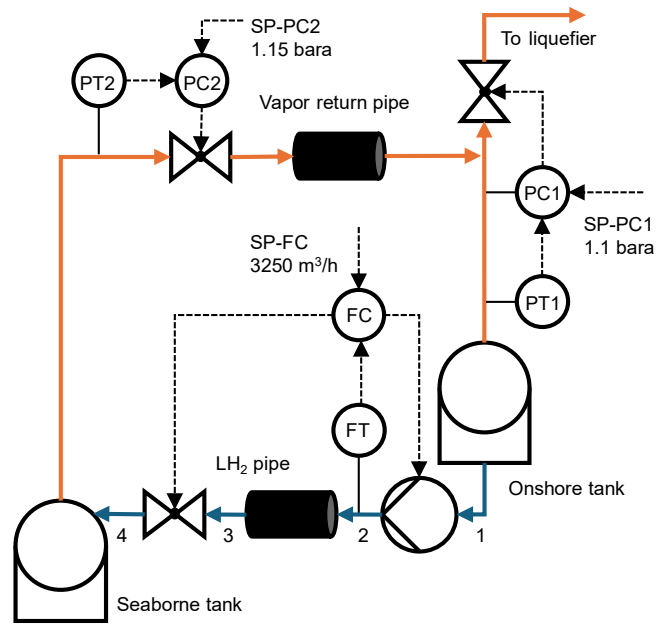

Figure 1: System with control structure and numbered state points (1-4) at the LH$_2$ transfer line. FT is a flow transmitter, FC flow controller and SP-FC its set-point. Similar for pressure (PT, PC, SP-PC). Blue lines indicate LH$_2$, orange lines gaseous H$_2$. The figure is simplified, as in reality the pump is submerged in the onshore tank.

### 2.2   LH$_2$ pump

We assume a centrifugal pump described by 2$^{nd}$ order characteristics with the possibility for a VSD. The best operating point is an efficiency $\eta = 60\%$ for a flowrate of 3250 m$^3$/h, providing 2 bar pressure increase at 60 Hz. Motor efficiency is not included. The VSD permits to reduce the speed to 25 Hz. The pump curves are shown in Figure 2.

Note that LH$_2$ pumps of this capacity do not exist yet. As an indication, the maximum throughput of LH$_2$ transfer pumps is 170 m$^3$/h in the Hydrogen Delivery Scenario Analysis Model (HDSAM) v 5.0 with an isentropic efficiency of 60% [12]. However, technology development is ongoing, and we assume in this work that centrifugal LH$_2$ pumps of high capacity will be available.

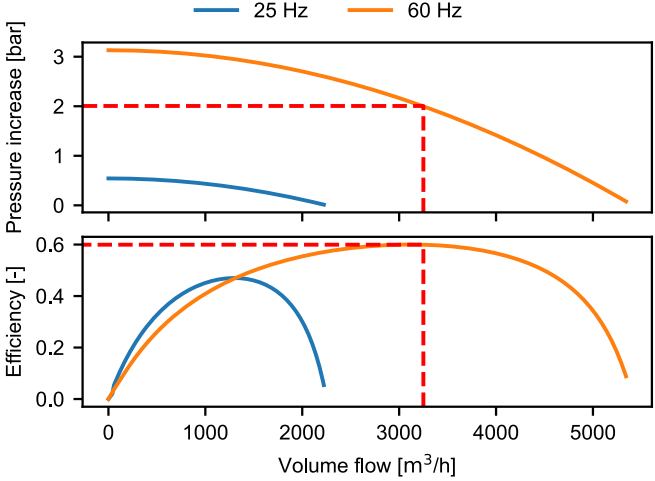

Figure 2: Operating range of the pump with VSD. The fixed-speed pump uses only the 60 Hz curve. Dashed line shows the flowrate for the nominal case.

### 2.3 Pipelines

Each train has two LH$_2$ pipelines and two vapor return lines. The pipelines are specified in Table 1 and modelled as horizontal pipes. Each pipe is discretized in 20 cells.

Table 1: Nominal pipeline values. Each train has 2 of these pipes.

|  | LH$_2$ pipe | BOG pipe |
|---|---|---|
| Internal diameter | 16'' (406 mm) | 18'' |
| Length | 1100 m | 1100 m |
| Heat ingress | 8.5 W/m | 1 W/m |
| Pipe roughness | 0.07 mm | 0.07 mm |

We apply the pressure drop model of Swamee-Jain which calculates the friction factor for the Darcy-Weisbach equation using the pipe roughness. Pipe roughness for new steel are 0.04 mm-0.1 mm, motivating our choice in Table 1 [9]. Values for the heat ingress to the LH$_2$ pipe is based on extrapolated values from datasheets of industrial suppliers [13], [14]. The alternative approach would be to model the insulation of the pipes.

Initial conditions of the pipelines are set as follows. The wall temperature is set to 20 K. The fluid enthalpy is set to the enthalpy during steady-state filling, which is found by an iterative approach.

Thermal expansion loops are not directly modelled. However, thermal expansion loops can be modelled by the concept of equivalent lengths [15]. Therefore, one may argue that our modelled pipes are shorter than stated in Table 1 but expansion loops are included. Furthermore, some companies develop cryogenic pipes which do not require expansion loops at all [16].

### 2.4 Seaborne tank and onshore tank

The tanks are modelled as separators in TIL with the parameters in Table 2. We stop filling the seaborne tank when it is 90% full, resulting in a total filling of 40 000 m$^3$.

Table 2: Storage tank parameters.

| Tank | Tank size | Initial fill | BOR |
|---|---|---|---|
| Onshore | 50000 m$^3$ | 90 vol% | 0.05%/day (8 kW) |
| Seaborne | 45000 m$^3$ | 1.1 vol% | 0.05%/day (7.5 kW) |

The BOR values in Table 2 are justified in Section 2.6. Furthermore, vapor-liquid equilibrium (VLE) is assumed in the tanks. The validity of this assumption is discussed in Section 4.1.

### 2.5 Thermodynamical analysis of the losses in the transfer process

The exergy destruction $\Delta E_d$ of a process defines the loss of theoretical useful work. An efficient transfer process must therefore minimize the exergy destruction, which is equivalent to minimizing the entropy production $\Delta S$ due to the Gouy-Stodola theorem:

$$\Delta E_d = T_0 \Delta S. \qquad (1)$$

Here, $T_0$ is the reference temperature of the environment. Therefore, we can determine the influence of each step in the transfer process from the OT to the ST in Figure 1 since

$$\Delta S_{OT \to ST} = \Delta S_{pump} + \Delta S_{LH_2\ pipe} + \Delta S_{valve}. \qquad (2)$$

### 2.6 Uncertainty and global sensitivity analysis for a fixed pressure in the seaborne tank

We perform an U&GSA for the system with VSD and split-range control. The U&GSA determines which uncertain input parameters in Table 3 drives the observed uncertainty in two selected key performance indicators (KPIs) [17]. The KPIs are defined in Section 2.6.1.

Table 3: Uncertain parameters for the U&GSA. All parameters are uncorrelated with a uniform probability distribution.

| Parameter | Low | High |
|---|---|---|
| Pump's efficiency [%] | 50 | 70 |
| Heat ingress, LH$_2$ pipe [W/m/pipe] | 5.5 | 12 |
| Pipe roughness [mm] | 0.04 | 0.15 |
| Set-point, flow control [m$^3$/h] | 2560 | 3585 |
| BOR onshore tank [%/day] | 0.045 | 0.123 |
| BOR seaborne tank [%/day] | 0.046 | 0.127 |

The low and high bounds in Table 3 were selected for the following reasons. Typical pump efficiencies were assumed [18, p. 261] since LH$_2$ pumps of this size do not exist yet. This fits well with the 60% efficiency for the (smaller) LH$_2$ transfer pump in HDSAM v 5.0 [12]. The pipe roughness for new steel is 0.04-0.1 mm and moderately encrusted steel 0.15 mm [9]. The set-point range for flow control ensures that the transfer operation is finished in reasonable time (10-16 hours). We calculated the BOR for each tank using the simple heat transfer method described in the Supporting Information, where we selected low and high limits for the overall heat transfer coefficient $U$ based on the literature survey in [19]. The BOR limits for the tanks differ since the tank sizes are different.

We apply the delta-moment independent measure method [20], [21] to calculate the first-order and $\delta$ sensitivity indices of the KPIs using SALib's implementation [22]. We selected the $\delta$-method in this case study over variance-based methods since we expect a non-symmetrical distribution for the KPI "Relative BOG" (defined in the next section), given that it is lower bounded by 0%. We used $N = 6000$ samples from a Latin hypercube. Confidence bounds of 95% for the sensitivity indices based on bootstrapping are provided.

### 2.6.1 Key performance indicators (outputs)

The first KPI is the "Relative BOG" when filling one tank with 40 000 m$^3$ (2812 tonnes) LH2. The Relative BOG [wt%] is given by

$$Relative\ BOG = \frac{\int \hat{m}^{BOG}(t)dt}{2812\ tonnes} \times 100, \quad (3)$$

where $\hat{m}^{BOG}$ is the flowrate from the onshore tank to the liquefier in Figure 1.

The second KPI is the "Relative power" [kJ/m$^3$], defined as the ratio of accumulated shaft power of the pump to 40000 m$^3$ LH$_2$

$$Relative\ power = \frac{Accumulated\ shaft\ power\ kJ}{40\ 000\ m^3}. \quad (4)$$

### 2.7 The influence of the maximum working pressure in the seaborne tank

We do a separate one-parameter-at-the-time sensitivity analysis for the maximum working pressure in the seaborne tank since it has a dominating influence on our KPIs. In this case study, we vary the maximum working pressure in the seaborne tank between 1.15 bara to 1.35 bara.

The benefit of a higher maximum working pressure is that there is more gas in the ullage, which is beneficial for the Relative BOG. However, there are some drawbacks. First, higher pressure may imply that the tank walls must be thicker (heavier and more expensive tanks). Second, the quality of LH$_2$ will be lower from the ship's perspective. The BOG problem is moved from an LH$_2$ terminal which has a large liquefaction plant, to a vessel which may have limited BOG handling capacities.

## 3 RESULTS

### 3.1 Fixed pressure in the seaborne tank

#### 3.1.1 Sensitivity of filling rate

Figure 3 shows the BOG amounts going to the liquefier for three different filling rates (SP-FC). Using split-range control significantly decreases the BOG flowrate to the liquefier, and we describe the reasons in detail in the next section. The erratic behavior of the BOG flowrate is due to the relatively tight control of the onshore tank pressure (PC1 in Figure 1). The tight control ensures there is enough pressure difference to transfer BOG from the seaborne tank to the onshore tank (the difference is only 0.05 bar if the pressures are at their set-point). Tight control means more usage of the control valve, resulting in a more erratic behavior of the flowrate.

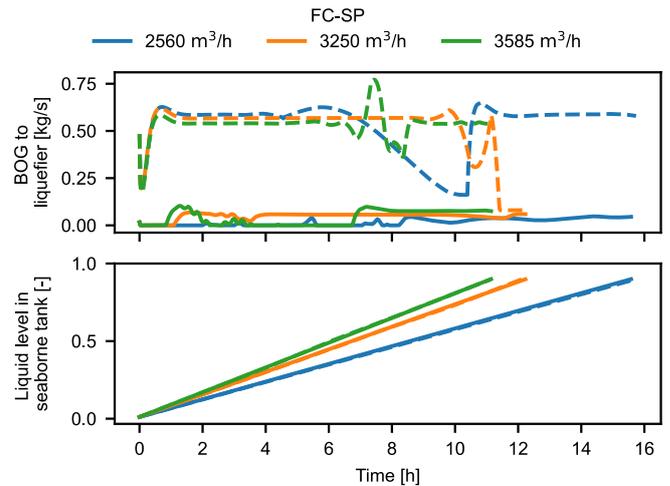

Figure 3: BOG to liquefier and filling times. Solid lines are for split-range control, dashed lines fixed-speed pump.

Figure 4 shows the KPIs defined by (3) and (4) for various filling rates (including the three filling rates in Figure 3). For the fixed speed pump, it is beneficial to have a high flowrate to minimize the accumulated BOG and power consumption. This is since a high flowrate implies a lower discharge pressure from the pump (Figure 2), and the next section describes why this is beneficial for the transfer process.

For the split-range controller, the slowest filling rate generally minimizes the Relative BOG. The exception is the last point with FC-SP = 3585 m$^3$/h. The more comprehensive U&GSA in Section 3.1.3 shows that the general trend is that slow filling gives lowest Relative BOG (see Figure 8). Furthermore, the slowest filling always gives the lowest power consumption.

These results show that the filling rate's influence on the Relative BOG depends on the pump type and operation. As stated in the Introduction, this nuances the findings in [6] where they conclude that a slow filling minimizes BOG formation (under the assumption of an idealized pump).

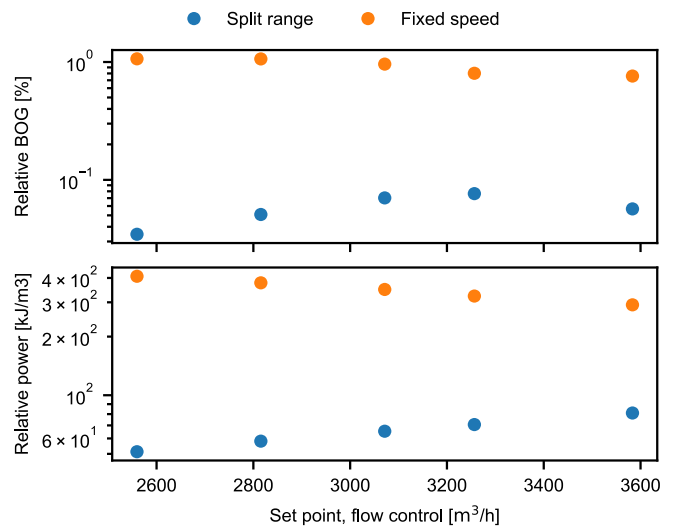

Figure 4: KPIs for various filling rates.

Table 4 shows the min/max values of the total BOG amount and pump's shaft power for our case of 160 000 m$^3$ LH$_2$ divided into four tanks.

Table 4: BOG amounts and total shaft power for the pump for the case of 160 000 m$^3$ LH$_2$ with nominal parameters.

| Flow controller | Total BOG | Total shaft power |
| --- | --- | --- |
| Split-range | 3.9-8.6 tonnes | 2.3-3.6 MWh |
| Fixed speed | 85.5-119.7 tonnes | 12.9-18.1 MWh |

*3.1.2 Thermodynamical analysis of the transfer process for fixed-speed pump and split-range control*

Figure 5 shows the TS-diagram at steady-state loading conditions when the flow controller's set-point is 3250 m$^3$/h for the two pump configurations. We observe that the entropy increase (proportional with exergy destruction) for the fixed-speed configuration is largest over the pump (point 1-2) and the throttling valve (point 3-4), while the contribution of the pipe (point 2-3) is relatively minor. This means that the exergy loss can be significantly reduced by avoiding unnecessary pressurization with subsequent throttling/depressurization. Therefore, for a fixed-speed pump we expect that more BOG is formed when reducing the flowrate since that implies a higher discharge pressure (see the pump chart, Figure 2).

From the fixed-speed pump case, we conclude that an inefficient transfer (higher entropy production) occurs when the $\Delta p$ in the system is high. The pressure increase is due to the pump, and one may argue that the fixed speed pump specified in Figure 2 has a too high head. However, an export terminal must be able to handle all kinds of ships, and there can be significant differences in the flow-resistance between ships. Therefore, it may not be feasible to down-size the pump.

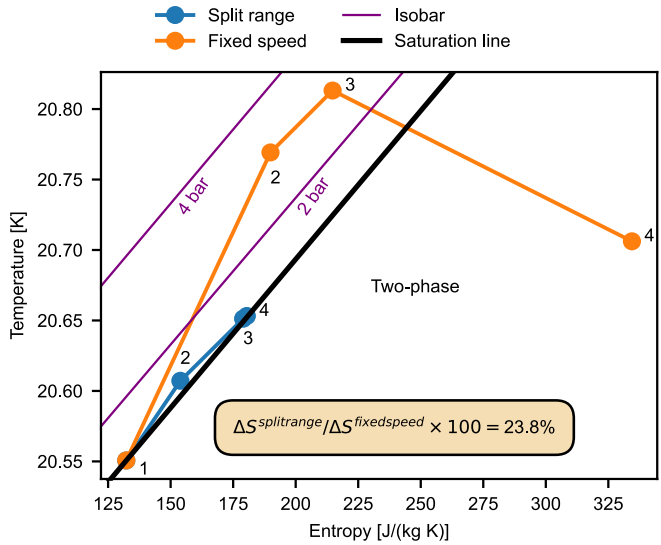

Figure 5: TS diagram. State points are numbered as in Figure 1. The entropy change for the split-range (SR) is calculated as $\Delta S^{SR} = S_4^{SR} - S_1^{SR}$, and similarly for the fixed speed.

In comparison, Figure 5 shows that the split-range controller excels and produces only 23.8% of the entropy compared to the fixed speed case. This is due to no throttling losses (points 3 and 4 are almost equal) and the pressure increase over the pump is smaller. We see that the entropy increase for the split-range case happens over the pump (points 1-2) and the pipe (points 2-3). The entropy increase for the pump can be reduced by increasing the efficiency of the pump, while for the pipe, it can be improved by reducing the pressure drop (e.g. its friction factor) and/or the heat ingress. Note that Figure 5 only considers one point in time (steady-state loading conditions), and the entropy production may vary during the bunkering process.

We conclude that minimizing the pressure increase along the transfer line is important. This is a complementary view to the findings in [6], [8], where only the receiving tank's pressure (point 4 in the TS diagram) were identified as a driver for Relative BOG.

*3.1.3 Uncertainty and sensitivity analysis*

Figure 6 shows the histograms for the two considered KPIs and statistics for the BOG flowrate to the liquefier (for one of four parallel trains). Interestingly, there are simulations where almost no BOG is going to the liquefier. This means that the volume of liquid leaving the onshore tank balances the expected pressure increase due to the vapor return from the seaborne tank. However, the gas phase in the onshore tank must be reliquefied once bunkering is completed and the tank is re-filled with LH$_2$ from the liquefier.

Statistics about BOG flowrate are important when designing the re-injection method of BOG to the liquefier. In Figure 6, the mean BOG flowrate to the liquefier never exceeds 0.19 kg/s, while only 17 of 6000 simulations have a max BOG flowrate exceeding 0.5 kg/s. The highest max-values may originate from controller tuning and initial conditions for valve openings which were "unfortunate" for the specific sample. In practical applications, where e.g. pump efficiency and desired filling rate (FC-SP) are better known, we may modify the controllers such that we do not see these extreme values.

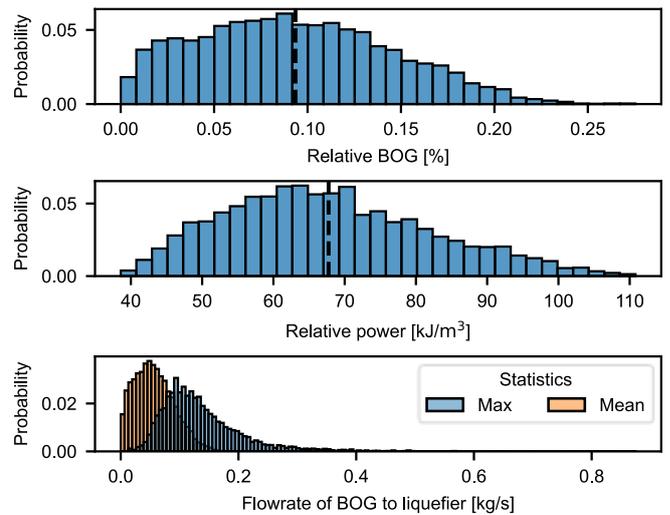

Figure 6: Histograms for the two KPIs and the flowrate of BOG to the liquefier (for one train). The mean of each KPI is marked with a black dashed line.

The relative BOG and relative power in Figure 6 are both skewed, which motivates our choice of using the delta moment-independent measure method for sensitivity analysis (a non-variance-based GSA method). Figure 7 shows the $\delta$-indices and first order sensitivity indices (S1) for all uncertain parameters. Clearly, the pump efficiency is the most important factor for both KPIs. Therefore, it is important to get a better understanding of the pump's efficiency to reduce the uncertainty in the KPIs (Figure 6).

The flow control set-point is the second most important factor for Relative power. For the Relative BOG, the heat ingress in $LH_2$ pipes is the second most important factor, although other factors are almost equally important. Finally, interaction effects between the parameters are important for Relative BOG since $\sum S1 = 64\%$, while this is not the case for Relative power ($\sum S1 = 100\%$).

Note that previous works [6], [8] have only identified the filling rate and pressure in receiving tank as important parameters. Our results clearly show why a GSA is called for, since we identify other important parameters which were previously overlooked. We stress that the results are subject to the specified uncertainties in Table 3 and assumptions of our model (which is different than the modelled systems in [6], [8]).

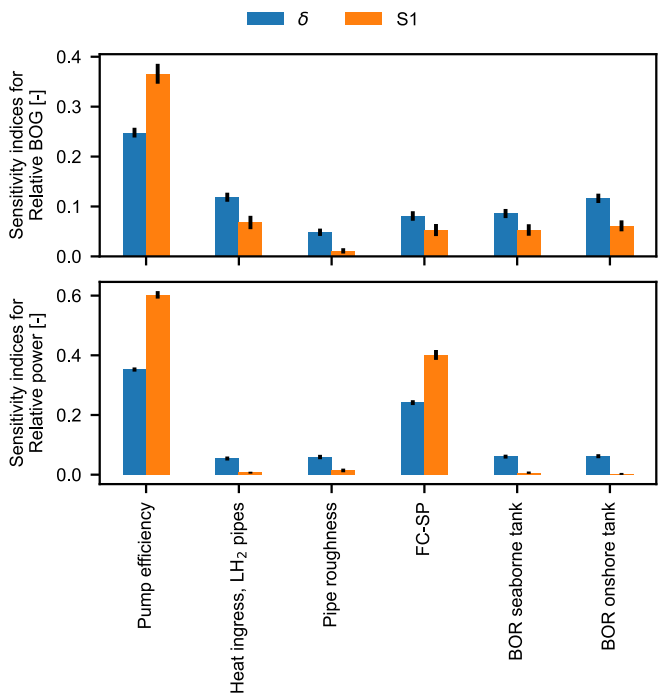

Figure 7: Sensitivity indices for the two KPIs.

The Relative BOG shown as a function of the filling rate and pump efficiency are shown in Figure 8. A slow filling rate and efficient pump minimizes the BOG formation when applying split-range control. The same holds for Relative power (not shown for brevity). The effect on Relative BOG for the remaining parameters in Table 3 is obvious (e.g. low heat ingress gives less BOG).

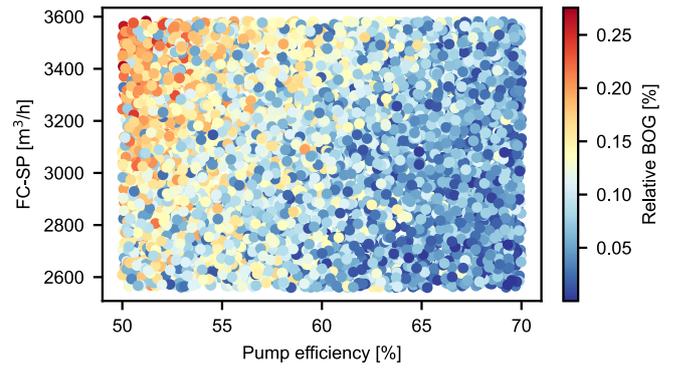

Figure 8: Influence of two important factors on the Relative BOG.

### 3.2 Sensitivity of pressure in seaborne tank

Figure 9 shows that the maximum working pressure in the seaborne tank has a dominating effect on our KPIs for the fixed-speed case. The Relative BOG decreases from 0.83% to zero when the maximum pressure is increased from 1.15 bara to 1.35 bara, respectively. Thus, a zero BOG transfer is also possible for the case of a fixed-speed pump if higher tank pressures are tolerated.

For the split-range case, the Relative BOG is zero when the seaborne tank's pressure is higher than 1.17 bara. Note that there is still vapor return from the ship to the onshore tank due to the displacement of ullage/cushion gas. However, this return flow rate is not sufficient to compensate for the amounts of $LH_2$ pumped out from the tank and thus maintain pressure. Therefore, the pressure in the onshore tank decreases below its set-point, and when the pressure in the seaborne-tank is higher than 1.22 bara the pressure in the onshore tank is below atmospheric. Note that the onshore tank does not receive $LH_2$ from the liquefaction plant in our simulations, which would to some extent counteract the decreasing pressure in the onshore tank.

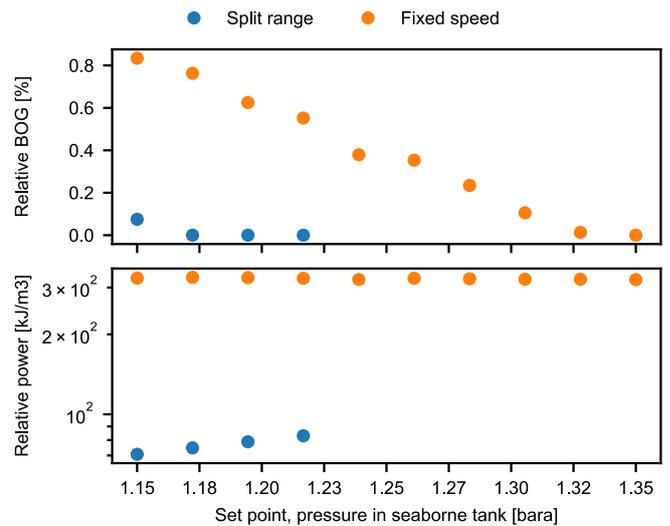

Figure 9: KPIs when varying the seaborne tank's maximum working pressure.

A drawback of increasing the tank pressure is that more BOG is stored in the seaborne tank when the pressure increases. This gives a worse starting point for the LH$_2$ carrier's BOG management system during the laden voyage. If the LH$_2$ carrier cannot consume all the BOG for propulsion, the BOG must be handled by e.g. an onboard reliquefaction system or flared to keep the tank pressure within the safe limit. Thus, the BOG issue is potentially moved from the export terminal (which *has* (re)liquefaction capabilities) to each vessel (which may not be able to reliquefy, and thus must flare or vent the BOG).

## 4 DISCUSSION

### 4.1 Storage tank model

The storage tanks are modelled as separators in TIL assuming VLE. This is likely a fair assumption for the seaborne tank since LH$_2$ is typically spray-injected from the top, giving a large contact surface and time between the gas and injected liquid. For the onshore tank, the VLE-assumption is more questionable since the gas phase is stratified. An improved model should incorporate this effect, alternatively model the non-equilibrium conditions between the liquid and gas phase as Petitpas [8], who slightly modified the work of Osipov [23].

### 4.2 Standby mode and BOG return to the liquefier

Our results indicate that it is possible to have zero loss during the transfer operation itself. However, we have not investigated how the infrastructure at the export terminal is affected by the time between ships (standby mode).

In the standby mode, we anticipate that the pipelines are kept cold by recirculating a small flow. This requires a bypass line between LH$_2$ pipes and vapor return pipes in Figure 1. However, the recirculation flow rate is likely significantly smaller than the flow rate during filling. Therefore, we expect that the pipes will warm up to some extent, and warmer pipes implies more BOG.

The pipe temperatures at the end of the standby mode correspond to the "correct" initial conditions for the pipes in our bunkering model. However, finding these initial conditions requires a detailed simulation of the standby mode and case-specific information. The pipe temperatures are affected by the duration of the standby mode, as well as the recirculation flow rate (see our paper for LH$_2$ bunkering systems for aviation [24]). Furthermore, the filling rate of LH$_2$ from the liquefaction plant to the onshore tank should be included in the analysis. The filling rate will affect the BOG rate from the onshore tank to the liquefaction plant by the "piston" effect (less gas volume is available when the liquid level raises).

Therefore, BOG is returned to the liquefier in two operation modes, during standby mode and potentially during bunkering. Our results indicate that zero or very low BOG flow during bunkering is feasible. In standby mode, we argued that the piston effect will displace BOG from the onshore tank back to the liquefier. This mode is expected to provide a stable BOG return flow, making ejectors a promising technology for reinjecting BOG into the liquefier, as briefly mentioned in the Introduction. However, peak shaving of BOG return may still be required, e.g. by using a (small) compressor to feed BOG to the liquefier. This presents an interesting direction for future work.

### 4.3 Vapor return from seaborne tank to onshore tank

We have assumed free flow of BOG between the two storage tanks, motivated by avoiding dedicated BOG compressors/blowers to save equipment cost. However, the pressure difference between the tanks is in our nominal case only 0.05 bar. This means that a large pipe diameter of the vapor return pipes is required. Furthermore, if the distance between the tanks is sufficiently long, free flow may not be feasible with such a small pressure difference, or it might be better to reduce the pipe diameter and install BOG blowers from an economic point of view.

## 5 CONCLUSION

This paper has shown that a zero-loss large-scale transfer of LH$_2$ can be possible. There are two strategies to achieve this. Option one is to use split-range control to regulate the flowrate using the pump's speed and a throttling valve. To get zero BOG with this option, the pump must have a high efficiency (close to 70%). Option two is to increase the maximum allowed pressure in the seaborne tank. This has the benefit that a fixed-speed pump can also be used. However, it may be a drawback from the ship's perspective to start the laden voyage with a higher tank pressure.

If the seaborne tank's pressure is not allowed to increase, the fixed-speed centrifugal LH$_2$ pump with some "safety factor" for the pressure increase ($\approx 2$ bar) gives unacceptable amounts of BOG at an LH$_2$ export terminal. We showed in a thermodynamical analysis that the issue lies with the (unnecessary/excessive) pressurization and subsequent throttling losses. On the other hand, using split-range control for the LH$_2$ flowrate provides reasonable BOG amounts and the desired operational flexibility.

For the case of split-range control, we performed an uncertainty analysis and found that the BOG amount when filling 160 000 m$^3$ (corresponding to 11 281 tons) LH$_2$ ranges from zero tons to 31 tons. The pump's efficiency is the uncertain parameter which reduces the produced BOG and power consumption of the pump the most. LH$_2$ pumps of the considered scale do not exist yet, but our results highlight that pump manufacturers should aim for high efficiency when developing LH$_2$ pumps since that gives significant system-wide savings for the end users.

**Declaration of generative AI and AI-assisted technologies in the manuscript preparation process**

During the preparation of this work the authors used ChatGPT in order to shorten the text and improve the language. After using this tool/service, the authors reviewed and edited the content as needed and take full responsibility for the content of the published article.

**Funding**

This publication is based on results from the research project *LH$_2$ Pioneer — Ultra-insulated seaborne containment system for global LH$_2$ ship transport*, performed under the ENERGIX program of the Research Council of Norway. The authors acknowledge the following parties for financial support: Gassco, Equinor, Air Liquide, HD Korea Shipbuilding & Offshore Engineering, Moss Maritime and the Research Council of Norway (320233).

**Declaration of competing interest**

The authors declare that they have no known competing financial interests or personal relationships that could have appeared to influence the work reported in this paper.

**CRediT authorsip contribution statement**

**Halvor Aarnes Krog:** Conceptualization, methodology, formal analysis, software, visualization, writing – original draft. **David Berstad:** Conceptualization, methodology, formal analysis, writing – review & editing, funding acquisition, project administration, supervision.

## Supporting Information – BOR calculations

The heat ingress into a tank is given by

$$Q_{tank} = (T_{amb} - T_{LH_2})A_{surface}U, \qquad (5)$$

where we used an ambient temperature $T_{amb} = 298.15$ K and the fluid temperature $T_{LH_2} = 20.55$ K (saturation temperature at 1.1 bara). $A_{surface}$ is the surface area of the spherical tank and $U$ is the overall heat transfer coefficient. We varied $U$ between 0.004 W/(m² K) and 0.011 W/(m² K), which are reasonable values for LH2 tanks [19]. The BOR, defined as the evaporation rate in a full tank, is then given by

$$BOR = Q_{tank}/(\rho_{LH_2}V_{tank}(1-ullage)h^{vap}_{LH_2}), \qquad (6)$$

where the density is $\rho_{LH_2} = 70.505$ kg/m³, $V_{tank}$ is the tank's volume, ullage is 0.1 and the heat of vaporization is $h^{vap}_{LH_2} = 444.7$ kJ/kg. The SI unit of BOR is $s^{-1}$, but it is typically given as mass%/day.